\documentclass[conference]{IEEEtran}
\usepackage{url}
\usepackage{hyperref}

\usepackage{xspace}

\AtBeginDocument{%
  \providecommand\BibTeX{{%
    \normalfont B\kern-0.5em{\scshape i\kern-0.25em b}\kern-0.8em\TeX}}}

\usepackage{xspace}
\usepackage{tikz}

\usepackage{booktabs}
\usepackage[skins]{tcolorbox}
\usepackage{amsmath}
\usepackage{graphicx}
\usepackage{graphics}
\usepackage{subcaption}
\usepackage{makecell}
\usepackage{colortbl}
\definecolor{darkgreen}{rgb}{0.0, 0.3, 0.13}
\definecolor{darkred}{rgb}{0.2, 0.0, 0.13}
\definecolor{darkred2}{rgb}{0.7, 0.04, 0.04}
\definecolor{darkblue}{rgb}{0.04, 0.44, 0.7}

\usepackage{hyperref}

\usepackage{array}
\newcolumntype{L}[1]{>{\raggedright\let\newline\\\arraybackslash\hspace{0pt}}m{#1}}
\newcolumntype{C}[1]{>{\centering\let\newline\\\arraybackslash\hspace{0pt}}m{#1}}
\newcolumntype{R}[1]{>{\raggedleft\let\newline\\\arraybackslash\hspace{0pt}}m{#1}}

\usepackage{xspace}

\newcommand{\BfPara}[1]{{\noindent\bf #1.}\xspace}

\newcommand{\etal}{{\em et al.}\xspace}

\usepackage{color}
\usepackage{xcolor}
\usepackage{longtable}
\usepackage{listings}

\definecolor{lightred}{rgb}{1, 0.8, 0.8}
\definecolor{lightgreen}{rgb}{0.8, 1, 0.8}

\lstdefinestyle{myStyle}{
  belowcaptionskip=1\baselineskip,
  breaklines=true,
  columns=fullflexible,
  language=Java,
  showstringspaces=false,
  basicstyle=\scriptsize\ttfamily,
  keywordstyle=\bfseries\color{blue},
  commentstyle=\itshape\color{green!40!black},
  identifierstyle=\color{black},
  stringstyle=\color{red},
  numbers=left,
  firstnumber=1,
  escapeinside={(*@}{@*)} 
}

\usepackage{xcolor}
\usepackage{makecell}
\usepackage[flushleft]{threeparttable}
\usepackage{mdframed}
\usepackage{adjustbox}

\usepackage{pgfplots}
\usepackage{caption}
\usepackage{subcaption}
\usepackage{colortbl}
\usepackage{pgfplots}
\usepackage{pgfplotstable}
\usepackage{graphicx} 
\usepackage{booktabs} 
\usepackage{tabularx}
\usepackage{wasysym}
\usepackage{multirow}
\usepackage{array}    
\usepackage{xspace}
\usepackage{color}
\usepackage{xcolor}

\usepackage{blindtext}

\usepackage{etoolbox}
\makeatletter
\patchcmd{\maketitle}{\@copyrightspace}{}{}{}
\makeatother

\colorlet{lightgrey}{lightgray}
\usepackage{tikz}
\usepackage{pgf}
\usepackage{booktabs}
\usepackage{amsmath}
\usepackage{graphicx}
\usepackage{makecell}
\usepackage{hyperref}
\usepackage{xcolor}
\usepackage{tcolorbox}
\usepackage{colortbl}
\usepackage[utf8]{inputenc}
\usepackage{rotating}
\usepackage{adjustbox}
\usepackage{afterpage}
\usepackage{lscape}
\usepackage{pifont}

\tcbset{
  hypothesis/.style={
    colback=gray!10,
    colframe=black,
    fonttitle=\bfseries,
    boxrule=0.1mm,
    arc=0mm,
    outer arc=0mm,
    width=\columnwidth, 
    boxsep=1pt,
    before skip=3pt,
    after skip=3pt,
    floatplacement=H, 
  }
}

\usepackage{pgfplots}
\pgfplotsset{compat = newest}
\usetikzlibrary{patterns}

\tcbset{textmarker/.style={%
        enhanced,
        parbox=false,boxrule=0.1mm,boxsep=0mm,arc=0mm,
        outer arc=0mm,left=3mm,right=3mm,top=3pt,bottom=3pt,
        toptitle=3mm,bottomtitle=3mm}}
\newtcolorbox{hintBox}{textmarker,
    borderline west={6pt}{0pt}{yellow},
    colback=yellow!10!white}

\newtcolorbox{blueBox}{textmarker,
    borderline={0pt}{0pt}{blue!50!white},
    colback=blue!5!white}
    
\newtcolorbox{importantBox}{textmarker,
    borderline west={6pt}{0pt}{red},
    colback=red!10!white}
\newtcolorbox{noteBox}{textmarker,
    borderline west={6pt}{0pt}{green},
    colback=green!10!white}

\usepackage{colortbl}
\definecolor{darkgreen}{rgb}{0.0, 0.3, 0.13}
\definecolor{darkred}{rgb}{0.2, 0.0, 0.13}
\newcommand{\ccd}{\cellcolor{red!15!white}}

\usepackage{makecell}

\begin{document}
\sloppy

\newcommand{\jie}[1]{\textcolor{blue}{#1}}
\newcommand{\dm}[1]{\textcolor{red}{#1}}

\title{Evaluating Large Language Models in Vulnerability Detection Under Variable Context Windows}

\author{Jie Lin and David Mohaisen \\
{\em University of Central Florida}}

\maketitle 

\begin{abstract}
This study examines the impact of tokenized Java code length on the accuracy and explicitness of ten major LLMs in vulnerability detection. Using chi-square tests and known ground truth, we found inconsistencies across models: some, like GPT-4, Mistral, and Mixtral, showed robustness, while others exhibited a significant link between tokenized length and performance. We recommend future LLM development focus on minimizing the influence of input length for better vulnerability detection. Additionally, preprocessing techniques that reduce token count while preserving code structure could enhance LLM accuracy and explicitness in these tasks.
\end{abstract}

\section{Introduction}
\label{sec:introduction}
Software vulnerabilities are one of the most significant threats today, costing billion of dollars in damanges, and calling for a lot of efforts on their detection and mitigation using advances in machine and deep learning techniques~\cite{MohaisenAM15,AlasmaryKAPCAAN19,AlkinoonAAMSM24,AlthebeitiM23,AnwarACLM22,AbusnainaAAAJSN22}.  Recent research explored the application of Large Language Models (LLMs) in vulnerability detection. Thapa \etal~\cite{thapa2022transformerbased} developed a framework for identifying vulnerabilities in C/C++ source code, achieving superior performance. Similarly, Khare \etal~\cite{khare2023understanding} evaluated pre-trained LLMs in detecting security vulnerabilities in multiple settings, highlighting the potential of LLMs in vulnerability detection with crafted prompts.

The use of LLMs for vulnerability detection involves analyzing source code to identify vulnerabilities and is influenced by factors like architecture, training data, and tasks. An LLM should accurately identify vulnerabilities regardless of code length or complexity for effective detection. Sensitivity to tokenized code length can lead to inconsistent results, affecting reliability. To our knowledge, the impact of the context window on tokenized code length has not been explored despite its importance. Additionally, there is limited research on LLM performance with Java code, a gap this paper addresses.

This study investigates the relationship between tokenized code length and the accuracy of various LLMs in vulnerability detection. We used LLaMA2, CodeLLaMA, LLaMA3, Mistral, Mixtral, Gemma, CodeGemma, Phi-2, Phi-3, and GPT-4 to detect vulnerabilities in Java code files. Using chi-square tests, we evaluated how tokenized code length affects the accuracy and explicitness of LLM responses.

\BfPara{Contributions} 1) A comprehensive comparative analysis of ten major LLMs in vulnerability detection using Java source code, focusing on the relationship between tokenized input length and performance. 2) An evaluation of the influence of tokenized length on LLM response accuracy and explicitness, using chi-square tests. 3) The implementation of a unified tokenization strategy with Byte Pair Encoding (BPE) for consistent comparison across LLMs, enabling a fair analysis of the relationship between token count and responses.

\section{Related Work}
\label{sec:related_work}
The related work falls into two categories: understanding vulnerability detection performance and LLM development. We review both, noting that our focus is on popular LLM models, though the selection is not exhaustive.

\BfPara{Understanding} Several studies have examined LLM decisions. Karlsen \etal~\cite{karlsen2023benchmarking} benchmarked various LLMs for security analysis, emphasizing fine-tuning for domain adaptation. Dong \etal~\cite{dong2024exploring} explored positional information within and beyond LLMs' context window (CW), proposing training-free CW extension. Despite these efforts, no studies have addressed factors influencing the quality of LLM responses.

\BfPara{Models} Meta's LLaMA began with models from 7B to 65B parameters, with pre-normalization, SwiGLU activation functions, and rotary positional embeddings~\cite{TouvronLIMLLRGHARJGL23}. LLaMA 2 expanded the pretraining corpus and context length, incorporating grouped-query attention~\cite{TouvronMSAABBBBBB23}, and CodeLLaMA optimized for code generation, handling sequences up to 100,000 tokens~\cite{RoziereGGSGE23}. LLaMA 3 introduced enhanced tokenizers and improved long-context task performance~\cite{meta2024llama3}. 
Google's Gemma and CodeGemma used a decoder-only architecture with multi-query attention, RoPE embeddings, GeGLU activations, and RMSNorm, with strong performance in language understanding~\cite{gemmateam2024gemma, codegemma2024}. Mistral~\cite{JiangSMBCC23}, used grouped-query and sliding window attention for efficient inference. Mixtral used a Sparse Mixture of Experts (SMoE) approach~\cite{JiangSRSBCCH24}. Microsoft's Phi family, e.g., Phi-1 and Phi-1.5, emphasized data quality and scaling techniques~\cite{ li2023textbooks, abdin2023phi2, abdin2024phi3}.

The factors influencing the quality of LLM responses remain largely unexplored. Previous studies have explored various applications of LLMs, such as self-adaptation in software systems \cite{donakanti2024reimagining} and incremental processing of garden-path sentences \cite{li2024incremental}. However, none have specifically addressed whether tokenized input length impacts the accuracy and explicitness of LLM responses in vulnerability detection.

This study addresses this gap by examining the correlation between tokenized input code length and LLM response quality across various models and contributes to a deeper understanding of LLM functionality to inform the development of more effective and reliable vulnerability detection methods.

\section{Methodology}
\label{sec:methodology}

\subsection{Vulnerability Detection Pipeline}

Our dataset is derived from the Vul4J~\cite{BuiSF22}, which addresses reproducible Java vulnerabilities. Vul4J includes 79 Java vulnerabilities from 51 open-source projects, covering 25 Common Weakness Enumeration (CWE) types, and includes Proof of Vulnerability (PoV) tests, patches, and build information.

To enhance Vul4J, we implemented a data curation pipeline as follows. \ding{172} {\em Automated Data Retrieval}: Using the OpenCVE API, we automated the retrieval of vulnerability descriptions, enhancing the dataset with rich contextual information.
\ding{173} {\em Data Cleaning and Preprocessing}: The data underwent cleaning to remove unnecessary characters, ensuring quality and usability.
\ding{174} {\em Integration of Descriptive Data}: Cleaned CVE and CWE descriptions were integrated into the dataset, providing a richer context for each vulnerability.
\ding{175} {\em Source Code Retrieval}: We enriched the dataset with source code changes from GitHub repositories, using a custom script to extract pre-patched and post-patched versions.
\ding{176} {\em Cleaning Source Code Data}: Comments were removed from the source code to reduce context token length and avoid variability in LLM decisions.
\ding{177} {\em Manual Inspection and Exclusion}: Non-relevant files were excluded, ensuring the dataset remained focused and relevant to our research objectives. The final curated dataset includes 140 Java files corresponding to 74 unique vulnerabilities, each linked to a specific CVE ID, and on average, fixing a vulnerability required changes to approximately 1.89 files. 

\subsection{Model Selection}
\label{subsec:model_selection}

\begin{table}[t]
\centering
\renewcommand{\arraystretch}{0.98}
\caption{Model selection. $^1$: the mini version of the model.}\label{tab:model_selection}\vspace{-2mm}
\scalebox{0.99}{\begin{tabular}{|l|c|c|c|c|c|}
\hline
\textbf{MF} & \textbf{Param} & \textbf{Ver} & \textbf{Type}                & \textbf{Quant}  & \textbf{CW}         \\ \hline
LLaMA2               & 7B                 & -               & Chat           & q5\_K\_M    & 4096                              \\ \hline
LLaMA2               & 13B                & -               & Chat           & q5\_K\_M    & 4096                              \\ \hline
LLaMA2               & 70B                & -               & Chat           & q5\_K\_M    & 4096                              \\ \hline
CodeLLaMA            & 7B                 & -               & Instruct        & q5\_K\_M   & 16384                              \\ \hline
CodeLLaMA            & 34B                & -               & Instruct        & q5\_K\_M   & 16384                              \\ \hline
CodeLLaMA            & 70B                & -               & Instruct       & q5\_K\_M    & 2048                              \\ \hline
LLaMA3               & 8B                 & -               & Instruct       & q5\_K\_M    & 8192                              \\ \hline
LLaMA3               & 70B                & -               & Instruct       & q5\_K\_M    & 8192                              \\ \hline
Mistral              & 7B                 & v0.2            & Instruct        & q5\_K\_M   & 32768                              \\ \hline
Mixtral              & 8x7B               & v0.1            & Instruct        & q5\_K\_M   & 32768                              \\ \hline
Gemma                & 2B                 & v1.1            & Instruct       & q5\_K\_M    & 8192                              \\ \hline
Gemma                & 2B                 & v1.1            & Instruct       & fp16        & 8192                             \\ \hline
Gemma                & 7B                 & v1.1            & Instruct       & q5\_K\_M    & 8192                              \\ \hline
Gemma                & 7B                 & v1.1            & Instruct       & fp16        & 8192                             \\ \hline
CodeGemma            & 7B                 & v1.1            & Instruct       & q5\_K\_M    & 8192                              \\ \hline
CodeGemma            & 7B                 & v1.1            & Instruct       & fp16        & 8192                             \\ \hline
Phi2                  & 2.7B               & v2              & Chat           & q5\_K\_M    & 2048                              \\ \hline
Phi2                  & 2.7B               & v2              & Chat           & fp16        & 2048                             \\ \hline
Phi3                 & 3.8B$^1$        & -              & Instruct          & q5\_K\_M    & 4096                              \\ \hline
Phi3                 & 3.8B$^1$        & -              & Instruct          & fp16        & 4096                             \\ \hline
GPT-4                & -                & -             & Chat            & -             & -                             \\ \hline
\end{tabular}}\vspace{-3mm}
\end{table}

We use LLaMA2, CodeLLaMA, LLaMA3, Mistral, Mixtral, Gemma, CodeGemma, Phi-2, Phi-3, and GPT-4. These models were selected for their architectural innovations, performance benchmarks, and relevance in current research, covering various architectures, parameter sizes, and training objectives (see Table~\ref{tab:model_selection}). Quantization reduces model weight precision to decrease memory usage and increase inference speed. Our experiments used Q5\_K\_M quantization, balancing performance and resourcing efficiency. Due to discrepancies, details about GPT-4's CW and quantization are excluded.

Our study aims to evaluate the robustness of various LLMs in detecting vulnerabilities in Java code files, explicitly examining how tokenized code length influences their decision. By selecting diverse models across different families, parameter sizes, and quantization methods, we investigate the correlation between tokenized code length and the accuracy and explicitness of LLM responses. This approach helps identify which LLMs can provide reliable and explicit vulnerability detection irrespective of input token length, highlighting their effectiveness under varying conditions.

\subsection{Experimental Pipeline}
\label{sec:experimental_pipeline}

A key component of our pipeline is the system prompt:

\addvspace{3pt}
\noindent\fbox{\begin{minipage}{24.4em}
\BfPara{Prompt} You are an expert Java programmer who can carefully analyze the provided Java code. The goal is to judge if the provided code is vulnerable or not. Your answer should be concise by saying yes or no to represent the code's type. If it is vulnerable, then yes; otherwise, no. Also, please explain concisely why you made the decision.
\end{minipage}}
\vspace{0.1em}

This prompt sets a clear context for each LLM, ensuring response consistency. The model is expected to respond with a simple response, mimicking the behavior of a human expert. 


We designed two experimental pipelines to evaluate LLMs' performance under different conditions. \ding{182} \textbf{Restricted Context Window Pipeline}: All LLMs are limited to a CW of 2048 tokens, with a maximum output token limit of 2048. The temperature is set to 0.5 to encourage precise and focused answers. \ding{183} \textbf{Extended Context Window Pipeline}: Each LLM utilizes its maximum described CW (e.g., CodeLLaMA 7B with 16384 tokens), while the maximum output tokens remain at 2048 and the temperature at 0.5.

\subsection{Response Categorization}
\label{subsec:response_categorization}
Our study examines the relationship between tokenized code length and LLM performance in identifying vulnerabilities in Java code files. All files contain vulnerabilities, so we focus on the explicitness and correctness of the responses. Due to the varied LLM responses, we manually examine all responses. Responses often lack standard grammar, making automatic assessment difficult. For example, a response like ``No, the code is vulnerable'' requires context to understand explicitness. We categorize responses as follows: Correct Response (1) explicitly states the code is vulnerable; Incorrect Response (0) explicitly states the code is not vulnerable; Irrelevant Response (-1) does not state the vulnerability status or is irrelevant.

\BfPara{Evaluation Metrics} We consider two key aspects: Accuracy, i.e., a correct response (1) indicates the LLM correctly detects vulnerability; Explicitness—an explicit response (1 or 0) shows the LLM follows the prompt without hallucination. Explicitness reflects the LLMs' ability to state the vulnerability status clearly. By examining the correlation between tokenized code length and these categorized responses through chi-square tests, we assess if input length affects LLM performance in providing accurate and explicit responses, thereby understanding how well LLMs yield their decisions.

\subsection{Unified Tokenization Strategy}
To calculate the tokenized length of code, we used a Byte Pair Encoding (BPE) tokenizer~\cite{Gage1994ANA} with a vocabulary size of 30,000 tokens. Each LLM receives the raw system prompt and code files without preprocessing or tokenization, allowing each model to use its internal tokenization methods. Our tokenization is separate from the LLMs' detection process and is used solely for analysis, not affecting their prediction accuracy or explicitness. Using a single tokenization method provides a uniform measure of token counts, enabling standardized analysis of the relationship between token count and LLM responses across all models.


\section{Null Hypotheses}
\label{sec:null_hypotheses}

We set two null hypotheses to statistically analyze the relationship between the tokenized length of input Java source code and the output of various LLMs in vulnerability detection. The choice of these null hypotheses is based on the notion that an optimal LLM should be able to provide accurate and explicit responses irrespective of the tokenized code length.

\BfPara{Null Hypothesis 1}
The first null hypothesis posits no relationship between the LLM's response indicating vulnerability and the tokenized length of the input Java source code. This suggests that the tokenized code length does not influence the LLM's ability to identify vulnerabilities correctly. Formally:

\addvspace{3pt}
\noindent\fbox{\begin{minipage}{24.4em}
\BfPara{H0\textsubscript{1}} No correlation between the LLM's response indicating the input source code is vulnerable and the tokenized length of the input Java source code.
\end{minipage}}\vspace{0.1em}

\BfPara{Null Hypothesis 2}
The second null hypothesis asserts no relationship between the LLM's explicitness in indicating whether the input source code is vulnerable or not and the tokenized length of the input Java source code. This implies that the explicitness of the response is not affected by the tokenized code length. Formally:


\addvspace{3pt}
\noindent\fbox{\begin{minipage}{24.4em}
\BfPara{H0\textsubscript{2}} No correlation between the LLM's response indicating whether the input source code is vulnerable or not vulnerable and the tokenized length of the input Java source code.
\end{minipage}}
\vspace{0.1em}

Accepting these null hypotheses means the LLM is not influenced by the input code length, which is a desirable characteristic for effective detection. Conversely, rejecting the null hypotheses would indicate that the tokenized code length plays a significant role in shaping the quality of LLM output, highlighting potential limitations and areas for improvement in the models' ability to handle diverse code inputs.

\BfPara{Chi-Square Test Settings}
\label{sec:chi-square-settings}
To evaluate the null hypotheses, we employ chi-square tests requiring parameter settings appropriately to ensure robust results. We use the following parameters:

\BfPara{\ding{71} Effect Size} The size is set to 0.3 (medium effect size, Cohen's $w$), allowing us to detect moderate associations between the input code length and LLM responses. 

\BfPara{\ding{71} Significance Level} Set to 0.05, indicating a 5\% risk of rejecting the null hypothesis when it is true. A p-value less than 0.05 will lead to rejecting the null hypothesis, indicating a statistically significant relationship. \textbf{Power:} Set to 0.80, aiming for an 80\% probability of correctly rejecting the null hypothesis when it is false, ensuring the test is sensitive enough to detect true effects.

\begin{table}[t]
\centering
\renewcommand{\arraystretch}{0.99}
\caption{Chi-square test results; accuracy (CHI\_A), explicitness (CHI\_E), rejected (R),  and accepted (A). Rows where both null hypotheses are accepted are highlighted.}
\label{tab:model_results}\vspace{-2mm}
\scalebox{0.99}{\begin{tabular}{|l|c|c|c|c|c|}
\hline
\textbf{Model Name} & \textbf{Param} & \textbf{Quant} & \textbf{CW} & \textbf{CHI\_A} & \textbf{CHI\_E} \\ \hline
LLaMA2              & 7B             & q5\_K\_M       & 2048        & R          & R              \\ \hline
LLaMA2              & 7B             & q5\_K\_M       & 4096        & R                  &  R                     \\ \hline
LLaMA2              & 13B            & q5\_K\_M       & 2048        & A          & R              \\ \hline
LLaMA2              & 13B            & q5\_K\_M       & 4096        &  R                 &    R                   \\ \hline
LLaMA2              & 70B            & q5\_K\_M       & 2048        & R          & R              \\ \hline
LLaMA2              & 70B            & q5\_K\_M       & 4096        &  R                 &    R                   \\ \hline
CodeLLaMA           & 7B             & q5\_K\_M       & 2048        & A          & R              \\ \hline
CodeLLaMA           & 7B             & q5\_K\_M       & 16384       &   A                &   R                    \\ \hline
CodeLLaMA           & 34B            & q5\_K\_M       & 2048        & A          & R              \\ \hline
CodeLLaMA           & 34B            & q5\_K\_M       & 16384       &   A                &   R                    \\ \hline
CodeLLaMA           & 70B            & q5\_K\_M       & 2048        & R          & R              \\ \hline
LLaMA3              & 8B             & q5\_K\_M       & 2048        & A          & R              \\ \hline
LLaMA3              & 8B             & q5\_K\_M       & 8192        &  R                 &    R                   \\ \hline
LLaMA3              & 70B            & q5\_K\_M       & 2048        & A          & R              \\ \hline
LLaMA3              & 70B            & q5\_K\_M       & 8192        &   A                &    R                   \\ \hline
Mistral             & 7B             & q5\_K\_M       & 2048        & A          & R              \\ \hline
\ccd Mistral             &\ccd 7B             &\ccd q5\_K\_M       &\ccd 32768       &\ccd   A                &\ccd     A                  \\ \hline
Mixtral             & 8x7B           & q5\_K\_M       & 2048        & A          & R              \\ \hline
\ccd Mixtral             &\ccd 8x7B           &\ccd q5\_K\_M       &\ccd 32768       &\ccd    A               &\ccd   A                    \\ \hline
Gemma               & 2B             & q5\_K\_M       & 2048        & R          & R              \\ \hline
Gemma               & 2B             & q5\_K\_M       & 8192        &    R               &     R                  \\ \hline
Gemma               & 2B             & fp16           & 2048        & R          & R              \\ \hline
Gemma               & 2B             & fp16           & 8192        &     R              &     R                  \\ \hline
Gemma               & 7B             & q5\_K\_M       & 2048        & R          & R              \\ \hline
Gemma               & 7B             & q5\_K\_M       & 8192        &     R              &   R                    \\ \hline
Gemma               & 7B             & fp16           & 2048        & R          & R              \\ \hline
Gemma               & 7B             & fp16           & 8192        &      R             &     R                  \\ \hline
CodeGemma           & 7B             & q5\_K\_M       & 2048        & R          & R              \\ \hline
CodeGemma           & 7B             & q5\_K\_M       & 8192        &   R                &      R                 \\ \hline
CodeGemma           & 7B             & fp16           & 2048        & A          & R              \\ \hline
CodeGemma           & 7B             & fp16           & 8192        &    A               &     R                  \\ \hline
Phi2                & 2.7B           & q5\_K\_M       & 2048        & R          & R              \\ \hline
Phi2                & 2.7B           & fp16           & 2048        & R          & R              \\ \hline
Phi3                & 3.8B$^1$       & q5\_K\_M       & 2048        & R          & R              \\ \hline
Phi3                & 3.8B$^1$       & q5\_K\_M       & 4096        &   R                &    R                   \\ \hline
Phi3                & 3.8B$^1$       & fp16           & 2048        & A          & R              \\ \hline
Phi3                & 3.8B$^1$       & fp16           & 4096        &    R               &    R                   \\ \hline
\ccd GPT4                & \ccd -              &\ccd -              &\ccd -           &\ccd A          &\ccd A              \\ \hline
\end{tabular}}\vspace{-5mm}
\end{table}

\section{Results and Analysis}
\label{sec:results}

The chi-square tests for different LLM configurations, presented in Table~\ref{tab:model_results}, reveal significant patterns concerning model parameters, quantization methods, CW, and advancements in models. The results indicate how these factors influence the acceptance or rejection of the null hypotheses related to accuracy (H0\textsubscript{1}) and explicitness (H0\textsubscript{2}) of responses.

Acceptance of the null hypotheses suggests that the LLM's performance is robust and not influenced by the input code length, which is desirable for effective vulnerability detection tools. Conversely, rejection indicates that the tokenized code length significantly impacts the model's responses, highlighting potential limitations and areas for improvement.

For the {LLaMA2} family, an increase in parameters does not consistently lead to the acceptance of the null hypotheses. The 13B parameter model accepts the null hypothesis for accuracy (H0\textsubscript{1}) but rejects it for explicitness (H0\textsubscript{2}), whereas both the 7B and 70B parameter models reject both hypotheses. This indicates that simply increasing parameters does not guarantee robustness against tokenized length variations.

For {CodeLLaMA}, the 7B and 34B models accept the null hypothesis for accuracy (H0\textsubscript{1}) but reject it for explicitness (H0\textsubscript{2}), regardless of the CW. The 70B model rejects both, suggesting that mid-range parameter models perform better in accuracy, though explicitness remains challenging. 

For {LLaMA3}, the 8B and 70B models accept the null hypothesis for accuracy (H0\textsubscript{1}) with a 2048 CW. However, increasing the CW to 8192 tokens leads to rejecting the null hypothesis for the 8B model, while the 70B model maintains acceptance for accuracy. This suggests that larger parameter models with appropriate CWs can maintain accuracy but struggle with explicitness for this model family. 

{Mistral} shows that the 7B model with a 2048 token CW accepts the null hypothesis for accuracy (H0\textsubscript{1}) but rejects it for explicitness (H0\textsubscript{2}). Increasing the CW to 32768 tokens leads to the acceptance of both null hypotheses, indicating that a significant increase in CW can mitigate the influence of tokenized length on both accuracy and explicitness. Similarly, the {Mixtral} 8x7B model performs better with the larger CW.

For {Gemma}, increasing the CW to 8192 tokens does not improve the acceptance of the null hypotheses. These models consistently reject both hypotheses, indicating these models are significantly influenced by tokenized input length.

{CodeGemma}, a fine-tuned Gemma, shows that increasing the quantization precision from q5\_K\_M to fp16 improves the acceptance of the null hypothesis for accuracy (H0\textsubscript{1}) but not for explicitness (H0\textsubscript{2}). This suggests that higher precision can reduce the influence of tokenized length on accuracy but does not have the same effect on explicitness for this model family.

{Phi}, especially Phi-3, shows improved performance with fp16 precision and a 2048 CW, accepting the null hypothesis for accuracy (H0\textsubscript{1}). This demonstrates that Phi-3's improved training techniques and model architectures contribute to better handling of tokenized input lengths for accuracy.

Finally, {GPT-4} consistently accepts both null hypotheses, demonstrating no significant relationship between tokenized length and the quality of responses in both accuracy and explicitness. This highlights GPT-4's robustness across different testing scenarios compared to other models.

\BfPara{Summary} Increasing parameters alone does not ensure robustness. Some models benefit from larger CWs, although this effect is inconsistent. LLMs for code understanding and generation, or those with advanced architectures, handle tokenized input length more effectively, showing the importance of configuring LLMs for specific needs.

\section{Conclusion}
\label{sec:conclusion}

Our study shows that for LLMs like Mistral, Mixtral, and GPT-4, there is no significant relationship between tokenized input length and response quality when the null hypotheses are accepted. This implies that adjusting tokenized length is unnecessary, simplifying vulnerability detection. GPT-4, which consistently accepts the null hypotheses, demonstrates robustness, making it reliable for Java code vulnerability detection. In contrast, models rejecting the hypotheses show a link between tokenized length and performance, suggesting areas for improvement. Identifying models less affected by input length streamlines detection and enhances LLM reliability.



\end{document}